# EFFICIENT REVERSIBLE DATA HIDING ALGORITHMS BASED ON DUAL PREDICTION

Enas N. Jaara and Iyad F. Jafar

Computer Engineering Department, The University of Jordan, Amman, Jordan

## ABSTRACT

*In this paper, a new reversible data hiding (RDH) algorithm that is based on the concept of shifting of prediction error histograms is proposed. The algorithm extends the efficient modification of prediction errors (MPE) algorithm by incorporating two predictors and using one prediction error value for data embedding. The motivation behind using two predictors is driven by the fact that predictors have different prediction accuracy which is directly related to the embedding capacity and quality of the stego image. The key feature of the proposed algorithm lies in using two predictors without the need to communicate additional overhead with the stego image. Basically, the identification of the predictor that is used during embedding is done through a set of rules. The proposed algorithm is further extended to use two and three bins in the prediction errors histogram in order to increase the embedding capacity. Performance evaluation of the proposed algorithm and its extensions showed the advantage of using two predictors in boosting the embedding capacity while providing competitive quality for the stego image.*

## 1. INTRODUCTION

Data hiding is an important technology in the areas of information security and multimedia copyright protections as it allows the concealment of data within the digital media for copyright protection and data protection. Many schemes of data hiding have been proposed to address the issues and challenges related to hiding the data, such as embedding capacity, imperceptibility and reversibility.

In this technique, the data is supposed to be seamlessly hidden or embedded into a carrier or cover signal (audio, images, video) in way that makes it hard for unauthorized people to access it [1]. In the digital imaging domain, several data hiding techniques have been proposed [2-4]. Despite the efficiency of these techniques in protecting the data, most of them are not capable of restoring the original cover image upon the extraction of embedded data. This poses a challenge to applications that require the preservation of the cover image after the hidden data is extracted. Accordingly, a great interest has grown in the past few years in the development of reversible data hiding (RDH) techniques that are capable of restoring the original image.

Several RDH techniques have been proposed in the literature and they compete in different aspects which include the embedding capacity, the quality of the stego image, size of overhead information and computational complexity [2]. Generally, they can be grouped into three different classes based on the concept of operation: difference expansion, histogram shifting, and

                                                                                    1

prediction-based techniques. Difference expansion (DE) algorithms are one popular class of reversible data hiding that are characterized with low distortion and relatively high embedding capacity. The first difference expansion technique was proposed by Tian in [5]. In this technique, the cover image is partitioned into a series of non-overlapping pixel pairs. A secret bit is then embedded using the difference expansion of each pixel pair. Several DE-based algorithms were developed based on Tian's technique [6-9]. Alattar [6] used DE with vectors instead of pixel pairs to extend and improve the performance of Tian's algorithm. Hu, *et al.* proposed a DE-based technique that improved the compressibility of the location map [8]. Compared to traditional DE-based algorithm, their technique increased the embedding capacity and performed well with different images.

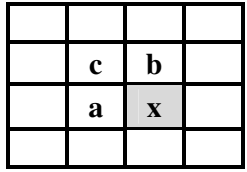

Figure 1 Context for predicting pixel x

Another important category of RDH algorithms are those that are based on the idea of histogram shifting (HS) [10-13]. Actually, the basis of these algorithms is the work presented by Ni, *et al.* [13]. In this algorithm, the histogram of the intensities in the original image is computed. Then, the histogram bins that lie between the peak bin and a zero (or minimum) bin is shifted by one toward the zero bin to open space to embedded data. Afterwards, the secret data bits are embedded by modifying the intensity value that corresponds to the peak only. This technique provided reasonable embedding capacity with minimum peak-signal-to-noise ratio (PSNR) of 48.1 dB. However, the main drawback of this technique is the limited hiding capacity due to the fact that it is dependent on the pixel count of the peak value, which is relatively low in natural images. Additionally, the embedded secret data cannot be recovered without knowing the values of peak and zero point of histogram. So the peak and zero points must be recorded as overhead or side information. Many algorithms were proposed to enhance the embedding capacity of Ni's algorithm while taking its advantage of producing high quality stego images. Hwang, *et al.* [10] extended Ni's algorithm by using two zero points and one peak point of the histogram to embed the data. Lin, *et al.* [12] employed multilevel hiding strategy to obtain high capacity and low distortion.

In order to take advantage of the HS techniques in terms of reversibility, several techniques attempted to overcome the issue of limited embedding capacity by extending the approach to histogram of prediction errors. Basically, these techniques modify the values of the prediction errors, which are computed using some predictor, instead of the actual intensities. The use of prediction errors is motivated by the fact that these errors are sharply centred near zero. This implies higher embedding capacities and avoids the need to save the peaks and zeros when compared to the original HS algorithm. Hong, *et al.* proposed extending Ni's algorithm by using the median edge detector (MED) [15]. The MED predicator computes the prediction $p$ of pixel $x$ using three neighbouring pixels a, b and c using

$$p = \begin{cases} \min(a, b), & \text{if } c \geq \max(a, b) \\ \max(a, b), & \text{if } c \leq \min(a, b) \\ a + b - c, & \text{othrwise} \end{cases} \qquad (1)$$

where a, b and c pixels are defined with respect to pixel x as shown in Figure 1. Afterwards, the prediction error (*PE*) which is the difference between pixel value and its prediction is computed. These prediction errors are changed based on their values and the bits of the secret message. Basically, the error values of 0 and -1 are used for embedding only. On other hand, prediction errors greater than 1 and less than -1 are incremented and decremented by 1, respectively. This is done to free the histogram bins at 1 and -2 to allow embedding of secret bits with value of 1, while zero bits are embedded in the 0 and -1 bins. The modified prediction errors are added to the prediction to produce the new values of the pixels in the stego image, the cover image after embedding the data. The algorithm showed remarkable results in terms of embedding capacity when compared to the original HS algorithm and it guaranteed a 48.1 dB as a lower bound for the quality of the stego image.

Several algorithms utilized the concept in prediction in data hiding [16-19]. Hong, *et al.* [16] proposed a reversible data hiding technique that is based on image interpolation and the detection of smooth and complex regions in the host images. Li, *et al.* [17] and Lin, *et al.* [18] introduced an information hiding scheme, with reversibility, based on pixel-value-ordering (PVO) and prediction-error expansion.

One of the main issues of prediction-based reversible data hiding algorithms is related to the type of the predictor that is used to compute the prediction errors. The accuracy of the predictor affects the embedding capacity and the quality of the stego image. So many predictors were used in different data hiding algorithms in the literature. However, most proposed algorithms rely on using a single predictor. The objective of this paper is to improve the efficiency of prediction-based reversible data hiding algorithms by designing an algorithm that employs two predictors to improve the prediction accuracy, thus the embedding capacity.

The proposed algorithm is based on the efficient modification of prediction errors (MPE) algorithm; however, it incorporates two predictors and uses only one bin of the prediction errors histogram for embedding the data, and it is referred to as 1-Bin MPE2. The1-Bin MPE2 algorithm is further extended to use more prediction errors in the embedding phase in order to increase the embedding capacity. These extensions are referred to by 2-Bin MPE2 and 3-Bin MPE2 algorithms. The performance evaluation of the proposed algorithm showed its ability to increase the embedding capacity with competitive image quality. Additionally, no overhead information is added to cope with the increase in the number of predictors. The rest of this paper is organized as follows. In Section 2, the proposed algorithm and its extensions are presented, followed by the experimental results in Section 3. Finally, the paper is concluded in Section 4.

## 2. THE PROPOSED ALGORITHM

As we mentioned earlier, most prediction-based RDH rely on the use of a single predictor to compute predictions that are used for data embedding. This may put a limitation on the prediction accuracy, since different predictors behave differently at the same pixel in the image [20,21] which consequently affects the embedding capacity and possibly the visual quality of the stego image.

In this paper, we propose a new histogram-shifting RDH algorithm that is based on the idea of employing two predictors to compute the predictions in order to take advantage of the different capabilities of different predictors. This is expected to increase the prediction accuracy, hence, increasing the embedding capacity. Additionally, and unlike the MPE algorithm [15], the proposed algorithm uses one bin of the prediction errors histogram for embedding the data. This algorithm, referred as the 1-Bin dual predictor MPE (MPE2), is further extended to use more prediction errors in the embedding phase in order to increase the embedding capacity. These extensions are referred to by 2-Bin MPE2 and 3-Bin MPE2.

## 2.1. Design Considerations

The proposed algorithm is essentially based on the modification of prediction errors (MPE) algorithm [15]. In other words, it is a histogram shifting technique that is applied to prediction errors. However, the proposed 1-Bin MPE2 algorithm uses two predictors instead of a single one and uses one bin from the prediction errors histogram for embedding. Specifically, the algorithm uses the error value of 0 for embedding as it is usually the most occurring value in the prediction errors histogram of natural images. Accordingly, when the cover image is scanned during the embedding phase, two predictors are used to compute the prediction at each pixel. When the prediction error of any of these predictors is 0, this predictor is selected for embedding. However, identifying the selected predictor at each pixel and guaranteeing reversibility during the extraction pose a challenge on the design of the proposed algorithm.

Consider the case for embedding a secret bit of value 1 in a pixel with intensity value of 105 using two predictors. If the predictions of these predictors are P1 = 106 and P2 = 105, then the prediction errors between the original pixel value and these predictions are PE1 = -1 and PE2 = 0, respectively. Next, the predictor with prediction error of 0 is used for embedding, and in this example it is the second predictor. Since the bit to be embedded is 1 and the prediction error is 0, the modified pixel value in the stego image is computed by shifting the prediction error PE2 to the right by 1, i.e. the modified prediction error PE2 becomes 1 and the pixel value in the stego image is 106 which is computed by adding the modified prediction error to prediction.

Now, in the extraction step, the same two predictors are used to compute the predictions in the stego image and they have the same prediction values P1 = 106 and P2 = 105. However, the prediction errors have been increased by 1 and become PE1 = 0 and PE2 = 1 due to embedding since the pixel value in the stego image was incremented by 1. Based on these prediction errors, the extraction procedure cannot identify the predictor that was originally used for embedding since we have prediction errors of 0 and 1 which according to MPE algorithm [15] correspond to embedding of a secret bit of 0 in prediction error 0 or embedding of a secret bit of 1 in prediction error 0, respectively. This implies that this approach for identifying the predictor might not be reversible in some cases.

A straight forward approach to overcome the issue of determining the used predictor during the extraction phase is to store the identity of the predictor at each pixel as an overhead that is communicated with the stego image to ensure reversibility. However, for an image of size M×N pixels, the size of the overhead will be (M-1)×(N-1) bits given that one bit is used to store the identities of the two predictors.

Based on the previous discussion, extending the original MPE algorithm to use two predictors without the need to increase the overhead size and to guarantee reversibility requires adopting a strategy for the selection of the predictor that is independent of the change in the prediction error values after embedding. As a general observation, the prediction errors produced by the two predictors should not be shifted or modified in all cases. Thus, we investigated all possible cases and put several rules for modifying the errors based on the relation between the values of the prediction errors from the two predictors.

The first rule considers the case when one prediction error is 0. In this case, the prediction error of the other predictor is checked. If it is positive and the bit to be embedded is 1, then both prediction errors are incremented by 1, i.e. PE1 and PE2 become PE1+1 and PE2+1. On the other hand, if the prediction error of the second predictor is negative and the bit to embed is 1, then both errors are decremented by 1. In both cases, no changes are made to the prediction errors when the bit to embed is 0 as this will preserve the values of the prediction errors.

However, a careful look at this first rule reveals that this approach is not reversible for some cases. Specifically, this happens when none of the predictors have zero prediction error. The following example shows one out of many cases that may make this approach irreversible. Consider the embedding of a secret bit in a pixel with intensity value of 107 using two predictors. If the predictions of these predictors are P1 = 106 and P2 = 108, then the prediction errors between the original pixel value and these predictions are PE1 = 1 and PE2 = -1, respectively. Thus, none of the predictors produces a prediction error of 0. So, the pixel value should be only shifted without any embedding. The question now is where to shift the pixel? Suppose that we shift the data to the right, the modified pixel value in the stego image is computed by shifting the prediction error to right PE1 = PE1+1=2, and adding it to the predicted value of the first predictor, i.e. the value of the pixel in the stego image is 108. Now, in the extraction step, the same two predictors are used to compute the predictions in the stego image and they have the prediction values P1 = 106 and P2 = 108. However, the prediction errors were increased by 1 and become PE1 = 2 and PE2 = 0, since the pixel value in the stego image was incremented by 1. These values indicate that there was embedding a bit of 0 at the pixel since one of the prediction errors is 0, which completely wrong.

Now suppose that we shift the data to the left, the modified pixel value in the stego image is computed by shifting the prediction error to left is PE2 =PE2-1= -2, and adding it to the predicted value of the second predictor, i.e. the value of the pixel in the stego image is 106. Now, in the extraction step, the same two predictors are used to compute the predictions in the stego image and they have the same prediction values P1 = 106 and P2 = 108. However, the prediction errors decrease by 1 and become PE1 = 0 and PE2 = -2, since the pixel value in the stego image was decremented by 1. Again, these values indicate that there was embedding a bit of 0 at the pixel since one of the prediction errors is 0, which completely wrong.

Accordingly, we added a second rule that is applied to consider modifying the pixel value according to the sign of the prediction errors. The two prediction errors must be either positive or negative in order to modify the corresponding pixel value, i.e. the prediction errors should be unipolar. If the two prediction errors are positive, then both prediction errors are incremented by 1, i.e. PE1 and PE2 become PE1+1 and PE2+1. On the other hand, if the two prediction errors are negative, then both errors are decremented by 1 to make this approach reversible.

Going back to the previous example, if the predictions of a pixel with intensity value of 107 are P1 = 106 and P2 = 108, then the prediction errors between the original pixel value and these predictions are PE1 = 1 and PE2 = -1, respectively. Thus, none of the predictors produces a prediction error of 0 and as stated in the second rule, for any bipolar prediction error values, the value of the corresponding pixel will remain unchanged.

In the extraction step, the same two predictors are used to compute the predictions in the stego image and they have the same prediction values P1 = 106 and P2 = 108. Then the prediction errors unchanged PE1 = 1 and PE2 = -1. These values indicate that there was no embedding or shifting at the pixel since the two prediction errors must be either positive or negative in order to modify the corresponding pixel value, which is completely right and makes this approach reversible.

Let's consider a case when the two prediction errors are positive, consider the embedding of a secret bit in a pixel with intensity value of 27 using two predictors. If the predictions of these predictors are P1 = 26 and P2 = 22, then the prediction errors between the original pixel value and these predictions are PE1 = 1 and PE2 = 5, respectively. Thus, none of the predictors produces a prediction error of 0. So, the pixel value should be incremented by 1, i.e. the value of the pixel in the stego image is 28.

Now, in the extraction step, the same two predictors are used to compute the predictions in the stego image and they have the same prediction values P1 = 26 and P2 = 22. Then the prediction errors become PE1 = 2 and PE2 = 6. These values indicate that there was no embedding and there is shifting at this pixel since the two prediction errors are positives. To recover the original pixel value, the pixel value should be decremented by 1, i.e. the value of the pixel in the recovered image is 27 which is the same as in the original image.

In summary, adopting these two rules in the 1-Bin MPE2 algorithm adds no overhead to the embedding process for the purpose of predictor identification and it works whether the secret bit is 0 or 1, and whether there is embedding or not since in all cases the prediction errors from all predictors are adjusted with the same value. It is expected that the proposed algorithm will be capable of increasing the embedding capacity due to the use of two predictors which increases the possibility of obtaining a 0 prediction error. Additionally, the quality of the stego images in the proposed algorithm is expected to be higher than the original MPE algorithm since pixels with bipolar errors are not modified. In the following two subsections, we present the steps of the embedding and extractions procedures in the proposed algorithm.

## 2.2. The Embedding Procedure for the 1-Bin MPE2 Algorithm

To embed secret messages $S$, let $CI$ be an 8-bit grayscale image with size $M \times N$ and $(i, j)$ be the pixel located on row $i$ and column $j$ in image $CI$, $1 \leq i \leq M$, $1 \leq j \leq N$. $SI$ is the stego image and the size is the same as $CI$. Note how the cover image is only used to initialize the first row and first column of the stego image. The embedding procedure of our algorithm involves calculating the prediction errors from the neighbourhood of a given pixel, and then embedding the message bits in the modified prediction errors. The detailed embedding steps are as follows:

*Input:*      An $n$-bit secret message $S$ and 8-bits $M \times N$ grayscale cover image $CI$.

*Output:*      A $M \times N$ stego image $SI$, end of embedding position $L$, and a data structure O containing the overhead information.

*Step 1.*      Prepare an empty 8-bit $M \times N$ matrix $SI$ to store the stego image and initialize the first row and first column with the values of the first row and first column of $CI$.

*Step 2.*      For the pixels in the range $1 < i \leq M$ and $1 < j \leq N$, scan the image in a raster scan order. If the $CI(i,j)$ is 0 or 255, then record the location of the pixel in $O$ and go to Step 11.

*Step 3.*      Compute the predictions $P1(i, j)$ and $P2(i, j)$. Where $P1(i,j)$ and $P2(i,j)$ are the predicted values calculated using two predictors.

*Step 4.*      Calculate the prediction errors $PE1(i,j)$ and $PE2(i,j)$ which are the difference between the original value $CI(i,j)$ and the predicted values $P1(i,j)$ and $P2(i,j)$, i.e., $PE1(i,j) = CI(i,j) - P1(i,j)$ and $PE2(i,j) = CI(i,j) - P2(i,j)$.

*Step 5.* In case *PE1(i,j)* is 0 and *PE2(i,j)* is greater or equal to 0, then keep *PE1(i,j)* unchanged if the secret bit B is 0, or modify *PE1(i,j)* to *PE1(i,j)*+1 if the secret bit *B* is 1. Compute the pixel value in the stego image by *SI(i,j) = P1(i,j) + PE1(i,j)*. Go to Step 11.

*Step 6.* In case *PE2(i,j)* is 0 and *PE1(i,j)* is greater than 0, then keep *PE2(i,j)* unchanged if the secret bit B is 0, or modify *PE2(i,j)* to *PE2(i,j)*+1 if the secret bit *B* is 1. Compute the pixel value in the stego image by *SI(i,j) = P2(i,j) + PE2(i,j)*. Go to Step 11.

*Step 7.* In case *PE1(i,j)* is 0 and *PE2(i,j)* is less than 0, then keep *PE1(i,j)* unchanged if the secret bit is 0, or modify *PE1(i,j)* to *PE1(i,j)*-1 if the secret bit is 1. Compute the pixel value in the stego image by *SI(i,j) = P1(i,j) + PE1(i,j)*. Go to Step 11.

*Step 8.* In case *PE2(i,j)* is 0 and *PE1(i,j)* is less than 0, then keep *PE2(i,j)* unchanged if the secret bit is 0, or modify *PE2(i,j)* to *PE2(i,j)*-1 if the secret bit is 1. Compute the pixel value in the stego image by *SI(i,j) = P2(i,j) + PE2(i,j)*. Go to Step 11.

*Step 9.* If both prediction errors *PE1(i,j)* and *PE2(i,j)* are greater than 0 or less than 0, modify *PE1(i,j)* to *PE1(i,j)*+1 or *PE1(i,j)*-1, respectively. Compute the pixel value in the stego image by *SI(i,j) = P1(i,j) + PE1(i,j)*. Go to Step 11.

*Step 10.* In case the prediction errors *PE1(i,j)* and *PE2(i,j)* are bipolar, then keep *SI(i,j)* unchanged.

*Step 11.* If all bits in the secret message *S* have been embedded, then record the location of the pixel *L* as the last embedding location. Go to Step 13.

*Step 12.* Update *i* and *j*. If pixel location (*M-1, N-1*) is not processed, then go to Step 2.

*Step 13.* Embedding is complete.

The assumption in this algorithm is that the used predictors are 3x3 and causal. Thus, the scanning in the embedding procedure excludes the first column and first row in the image, so, these pixels are not used for embedding. The algorithm can be easily modified to accommodate for larger and/or non causal predictors. Note that the overhead data structure *O* is used to save the locations of the pixels at which embedding may cause overflow or underflow. Because no changes are allowed to the prediction errors if the pixel value after modified is overflow or underflow case in order make this approach reversible. Fortunately, the size of overhead data structure *O* is often zero or negligibly small for most natural images since the overflow/underflow problem rarely occurs [15]. Also, the last embedding location in the embedding level is saved in the overhead data structure *O*. The flowchart of the embedding procedure is shown in Figure 2.

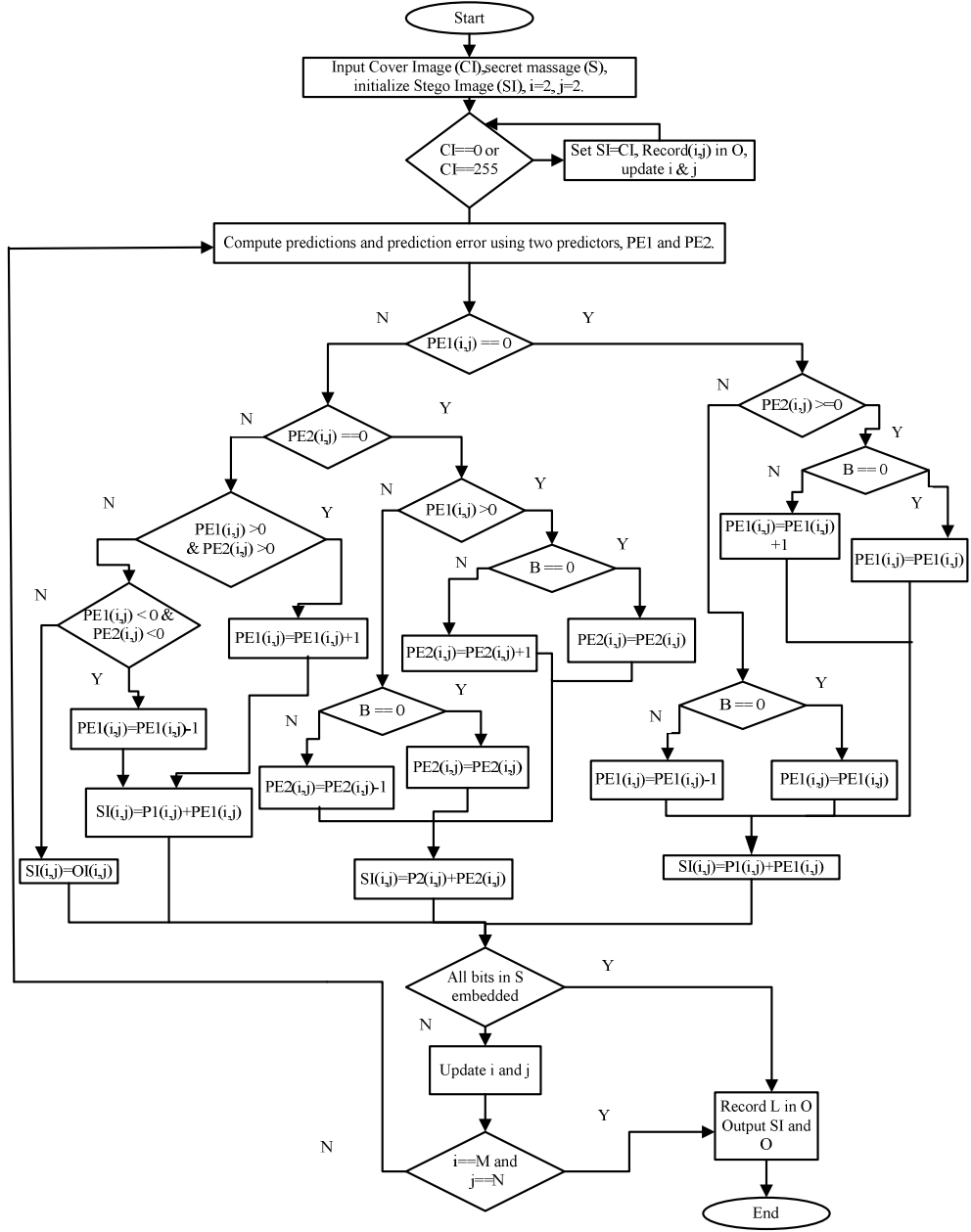

Figure 2. Flowchart of the embedding procedure of the 1-bin proposed algorithm

## 2.3. The Extraction Procedure of the 1-Bin MPE2 Algorithm

Using the same scan order sequence in the embedding procedure, we predict the pixel values again using the same predictors, and calculate the prediction errors *PE1* and *PE2* at each pixel in the stego image *SI*. Then we can restore the original image and the hidden data. The details of the extraction procedure are shown in Figure 3. The detailed steps for extracting hidden data and recovering the original image are as follows:

*Input:*    A $M{\times}N$ stego image *SI* and a data structure *O* containing the overhead information.

*Output:*    An *n*-bit secret message *S* and 8-bit $M{\times}N$ grayscale recovered image *OI*.

*Step 1.*    Prepare an empty 8-bit $M{\times}N$ matrix *OI* to store the output image and initialize the first row and first column with the values of the first row and first column of *SI*. Prepare an empty array to store the extracted secret message *S*.

*Step 2.*    For the pixels in the range $1 < i \le M$ and $1 < j \le N$, scan the image in a raster scan order. If the location of the pixel is found in *O*, then set *OI(i,j)= SI(i,j)*. Go to *Step 15*.

*Step 3.*    Compute the predictions *P1(i,j)* and *P2(i,j)*. Where *P1(i,j)* and *P2(i, j)* are the predicted values calculated using two predictors.

*Step 4.*    Calculate the prediction errors *PE1(i,j)* and *PE2(i,j)* which are the difference between the original value *SI(i,j)* and the predicted values *P1(i,j)* and *P2(i,j)*, i.e., *PE1(i,j) = SI(i,j) – P1(i,j)* and *PE2(i,j) = SI(i,j) – P2(i,j)*.

*Step 5.*    If *PE1(i,j)* is 0 and *PE2(i,j)* is greater or equal to 0, append a bit of 0 to *S* and keep *PE1(i,j)* unchanged. Compute the pixel value in the original image by *OI(i,j) = P1(i,j) + PE1(i,j)*. Go to Step 15.

*Step 6.*    If *PE1(i,j)* is 1 and *PE2(i,j)* is greater or equal to 1, append a bit of 1 to *S* and modify *PE1(i,j)* to *PE1-1*. Compute the pixel value in the original image by *OI(i,j) = P1(i,j) + PE1(i,j)*. Go to Step 15.

*Step 7.*    If *PE2(i,j)* is 0 and *PE1(i,j)* is greater than 0, append a bit of 0 to *S* and keep *PE2(i,j)* unchanged. Compute the pixel value in the original image by *OI(i,j) = P2(i,j) + PE2(i,j)*. Go to Step 15.

*Step 8.*    If *PE2(i,j)* is 1 and *PE1(i,j)* is greater than 1, append a bit of 1 to *S* and modify *PE2(i,j)* to *PE2-1*. Compute the pixel value in the original image by *OI(i,j) = P2(i,j) + PE2(i,j)*. Go to Step 15.

*Step 9.*    If *PE1(i,j)* is 0 and *PE2(i,j)* is less than 0, append a bit of 0 to *S* and keep *PE1(i,j)* unchanged. Compute the pixel value in the original image by *OI(i,j) = P1(i,j) + PE1(i,j)*. Go to Step 15.

*Step 10.*    If *PE1(i,j)* is -1 and *PE2(i,j)* is less -1, append a bit of 1 to *S* and modify *PE1(i,j)* to *PE1+1*. Compute the pixel value in the original image by *OI(i,j) = P1(i,j) + PE1(i,j)*. Go to Step 15.

*Step 11.*    If *PE2(i,j)* is 0 and *PE1(i,j)* is less than 0, append a bit of 0 to *S* and keep *PE2(i,j)* unchanged. Compute the pixel value in the original image by *OI(i,j) = P2(i,j) + PE2(i,j)*. Go to Step 15.

*Step 12.*    If *PE2(i,j)* is -1 and *PE1(i,j)* is less -1, append a bit of 1 to *S* and modify *PE2(i,j)* to *PE2+1*. Compute the pixel value in the original image by *OI(i,j) = P2(i,j) + PE2(i,j)*. Go to Step 15.

*Step 13.*    If both prediction errors, i.e. *PE1(i,j)* and *PE2(i,j)* are greater than 1 or less than -1, modify *PE1(i,j)* to *PE1(i,j)-1* or *PE1(i,j)+1*, respectively. Compute the pixel value in the original image by *OI(i,j) = P1(i,j) + PE1(i,j)*. Go to Step 15.

*Step 14.*    In case the prediction errors, i.e. *PE1(i,j)* and *PE2(i,j)* are bipolar, then keep *OI(i,j)* unchanged.

*Step 15.*    If the location of the pixel is *L*, extraction is complete. Otherwise, update *i* and *j* and go to Step 2.

*Step 16.*    Extraction is complete.

In order to demonstrate the idea of the proposed algorithm, we discuss here a detailed numerical example which considers all possible cases that might be encountered.  Suppose that we have a 4x5 image that is shown in Figure 4(a) and assuming the secret message *S* to be embedded is $(011101)_2$. Here, we assume the predictions at each pixel are computed using the MED and mean predictors. The MED predictor is the predictor that is used in MPE algorithm [15]. The mean or average of three neighboring pixels [22, 23] computes the prediction value P using

$$P = \lfloor (a + b + c)/3 \rfloor \tag{2}$$

where $\lfloor \bullet \rfloor$ is the floor operator. As presented in the embedding procedure, the first step is to find the prediction values P1 and P2 using the MED and Mean predictors, respectively. The algorithm starts at location (2,2) and computes the predictions P1and P2 to be 6 and 3, respectively. Accordingly, the prediction errors will be 2 and 5. Since none of the prediction errors is 0 and both values are positive, no embedding is performed. However, PE1, which is the minimum, is incremented by 1 and added to P1to compute the pixel value in the stego image SI(2,2) to be 9 as shown in Figure 4(b).

The algorithm proceed to location (2,3) and computes the predictions P1 and P2 to be 9 and 4, respectively. Accordingly, the prediction errors will be 0 and 5. Since one of the prediction errors is 0 and other value is positive, embedding can be performed. However, the first bit in the secret data is zero, so PE1 is unchanged and added to P1 to compute the pixel value in the stego image SI(2,3) to be 9 as shown in Figure 4(c).

For location (2,4), the predictions are 9 and 5, respectively. Accordingly, the prediction errors will be -5 and -1. Since none of the prediction errors is 0 and both are negative, no embedding is performed. However, PE1, which is the minimum, is decremented by 1 and added to P1 to compute the pixel value in the stego image SI(2,4) to be 3 as shown in Figure 4(d).

At location (2,5), the computed predictions P1 and P2 are 3 and 4, respectively. Accordingly, the prediction errors will be 0 and -1. Since one of the prediction errors is 0 and other value is negative, embedding can be performed. Since the second bit in the secret data is one, PE1, is decremented by 1 and added to P1 to computed the pixel value in the stego image SI(2,5) to be 2 as shown in Figure 4(e). After hiding the secret data 011101 in the original image, the stego image is shown in Figure 4(f).

In the extraction process, we use the same scan order as in the embedding process to predict pixel values again, and calculate the prediction errors PE1 and PE2 using the same two predictors. We realize that, when the value PE1 or PE2 is 0, then the embedded secret bit is 0. However, when the value PE1 or PE2 is 1 or -1, then a secret bit of value 1 is embedded in this pixel. Lastly, if the value PE1 or PE2 is not one of the three numbers -1, 0 and 1, then there is no bit embedded. Since we have changed the prediction errors during embedding, the original image can be recovered by modifying the prediction errors back to their original. Figure 5(a) shows the stego image. As shown in the extracting procedures, the first step is to find the predictive pixel values P1 and P2 for location (2,2). To compute the predictions P1and P2, we used the same MED and Mean predictors. P1 and P2 will have the same values obtained in the embedding procedure as shown in Figure 4(a). After extraction process ends, we obtain the secret data and the covered image again shown in Figure 5(f).

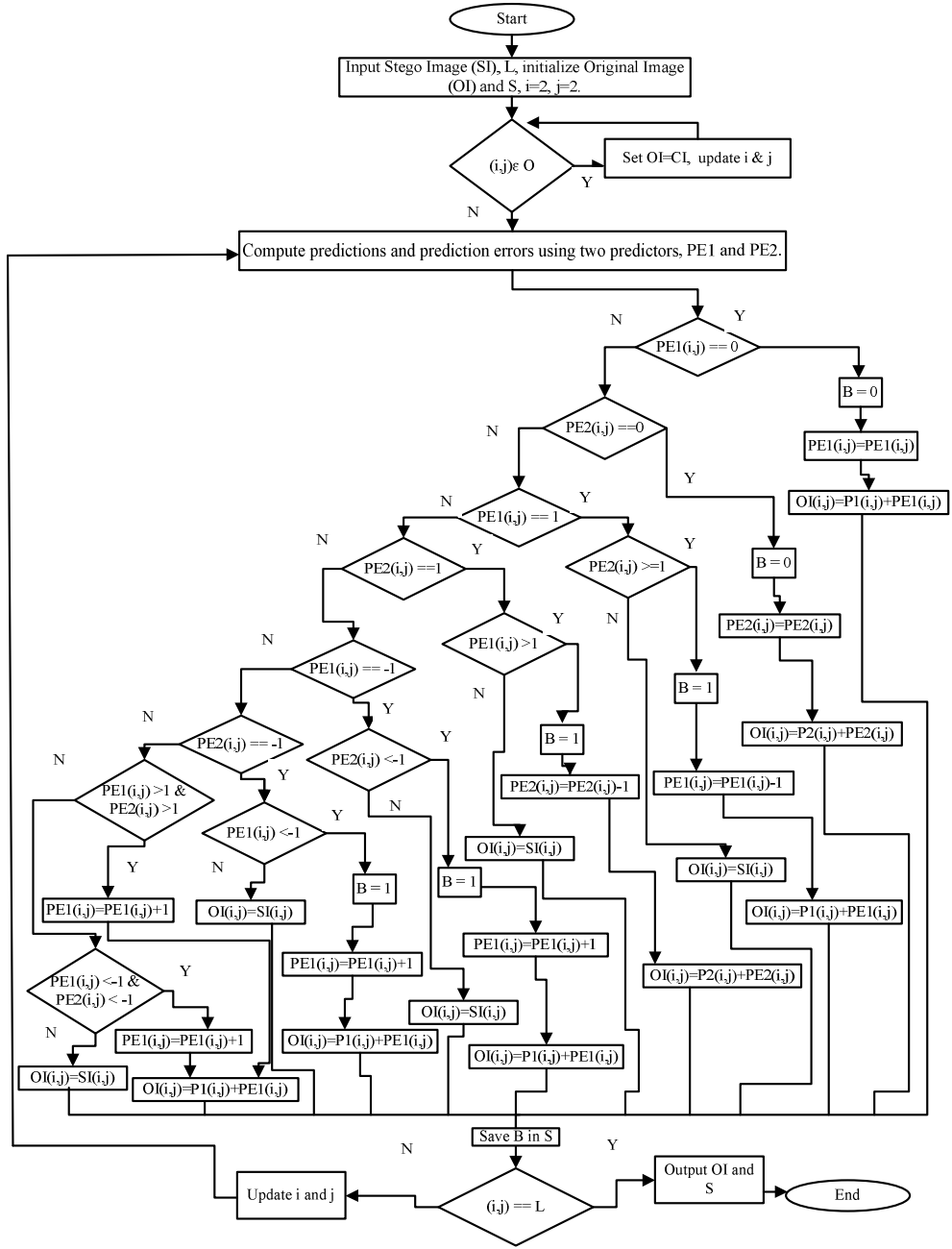

Figure 3. Flowchart of the extraction procedure of the 1-bin algorithm

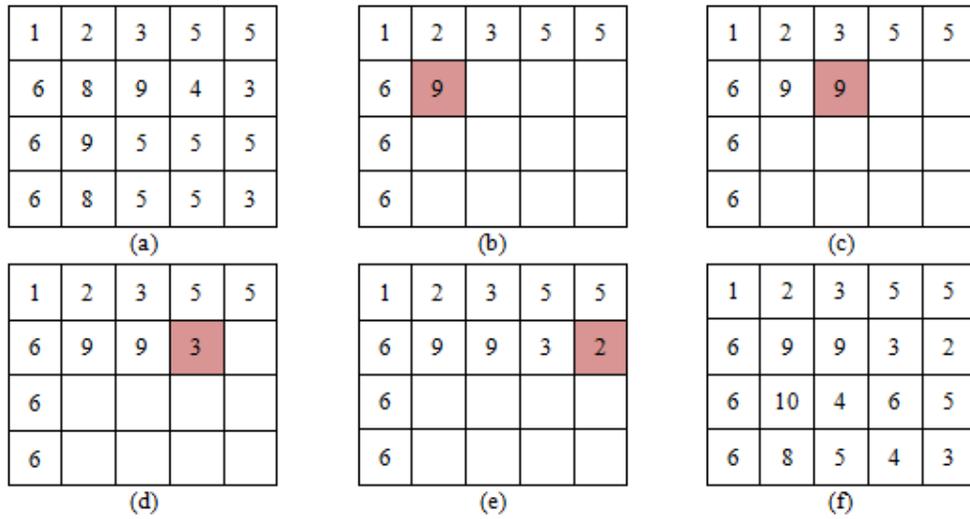

Figure 4. Embedding Example (a) Original image (b) – (e) embedding in consecutive pixels (f) stego image

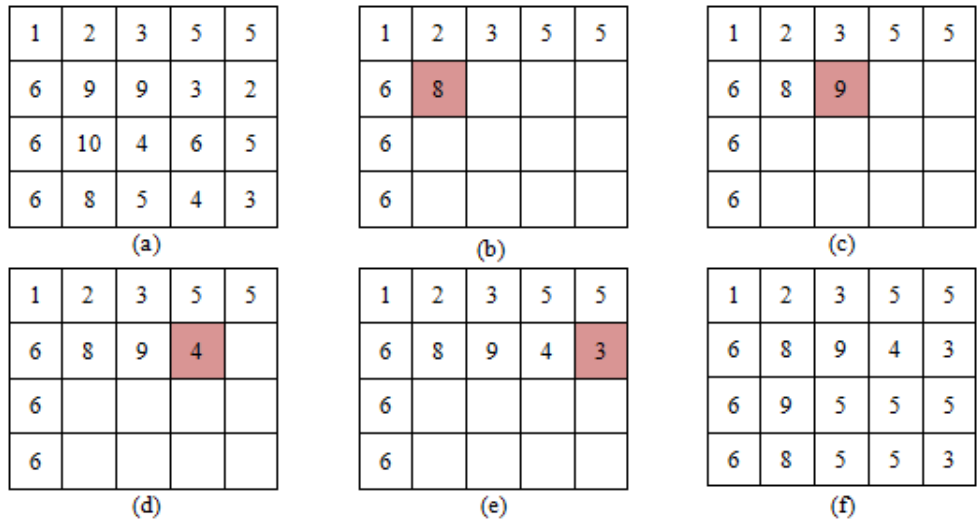

Figure 5. Extraction Example. Stego image is (a) while Images (b) through (e) show the extraction in consecutive pixels (f) the original image after extraction.

## 2.4. Extensions of the 1-Bin MPE2 Algorithm

The performance of the 1-Bin MPE2 algorithm can be further improved in terms of embedding capacity by using more prediction error values for embedding. In this section, we discuss the extension of the proposed algorithm to use two and three bins. We refer to these extensions as the 2-Bin MPE2 and 3-Bin MPE2 proposed algorithms. Also, we discuss extending the algorithm to incorporate more than two predictors.

A straight forward approach to increase the embedding capacity of the proposed 1-Bin MPE2 algorithm is to use more prediction error values for embedding. Here, we propose the 2-Bin MPE2 algorithm that uses error values of 0 and -1, which are the error values used by the 2-Bin

MPE algorithm. The steps in embedding procedures for the 2-Bin MPE2 algorithm are the essentially same as presented in the embedding procedure of the 1-Bin MPE2 algorithm. However, a third rule is added and used during the embedding and extraction phases. The new rule considers the case when one prediction error is -1, which is the new value that is considered for embedding. In this case, the prediction error of the other predictor is checked. If it is negative and the bit to embed is 0, then both prediction errors are decremented by 1, i.e. PE1 and PE2 become PE1-1 and PE2-1. And if the bit to embed is 1, then both prediction errors are decremented by 2, i.e. PE1 and PE2 become PE1-2 and PE2-2. On the other hand, if the prediction error of the second predictor is positive, no data will be embedded. Additionally, when the prediction errors of the two predictors are both negative and less than -1, they are shifted to the left by 2 instead of 1 to guarantee reversibility. However, this excessive shifting will result with images quality below 48.1 dB.

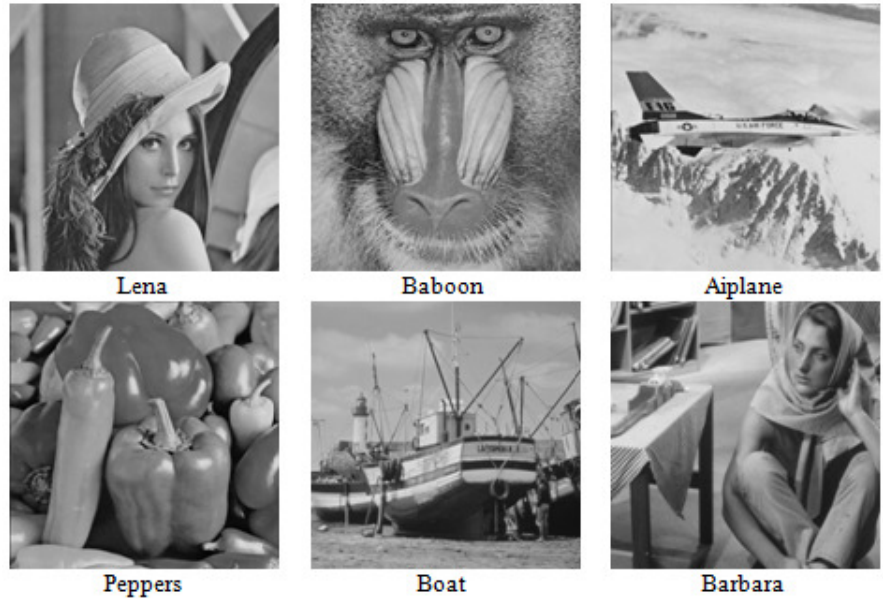

Figure 6. Benchmark test images.

A second simple extension to the 1-Bin MPE2 considers using error values of -1, 0 and 1 for embedding. Here, we propose the 3-Bin MPE2 algorithm that uses error values of 0, -1 and 1. The steps in embedding procedures for the 3-Bin MPE2 algorithm are the essentially same as presented in the embedding procedure of the 2-Bin MPE2 algorithm. However, a forth rule is now added and used during the embedding and extraction phases. The new rule considers the case when one prediction error is 1, which is the new value that is considered for embedding. In this case, the prediction error of the other predictor is checked. If it is positive and the bit to embed is 0, then both prediction errors are incremented by 1, i.e. PE1 and PE2 become PE1+1 and PE2+1. In case the bit to embed is 1, then both prediction errors are incremented by 2, i.e. PE1 and PE2 become PE1+2 and PE2+2. On the other hand, if the prediction error of the second predictor is negative, no data will be embedded. Additionally, when the prediction errors of the two predictors are both positive and greater than 1, they are shifted to the right by 2 instead of 1 to guarantee reversibility of this version. However, this excessive shifting will result with images quality below 48.1 dB.

# 3. EXPERIMENTAL RESULTS

In this section, we evaluate the performance of the proposed algorithm and its extensions and compare it with other RDH algorithms. The evaluation was performed on a large set of benchmark images that are usually used for this purpose. In this paper, we present sample results obtained for a set of commonly used benchmark images are obtained from (USC-SIPI Image Database) [24]. The test images are shown in Figure 6. In all experiments, a random secret message is generated an embedded in the images using different algorithms.

In the proposed algorithm and its extensions, the mean [23] and the MED [15] predictors are used due to their simplicity and efficiency. These predictors compute the predictions as given in (1) and (2). Nonetheless, this does not imply that the proposed algorithms work only with these predictors.

The proposed algorithm and its extensions are compared with the original version of the MPE algorithm [15], Hong, *et al.* algorithm [16], Li *et al.* [17] and Lin, *et al.* algorithm [18]. These algorithms are selected as they are prediction-based algorithms and they are proposed to produce high image quality at reasonable payloads. In addition, the comparison includes an extended version of the MPE algorithm. Specifically, this extension, 3-Bin MPE, uses three prediction errors (-1, 0, and 1) instead of two and is used to provide a fair comparison with the 2-Bin MPE2 algorithm.

The evaluation considered using two performance metrics; the peak signal-to-noise ratio (PSNR) and the pure embedded capacity (EC). The PSNR [25-27] of a stego image SI when compared with original cover image CI is given by

$$PSNR = 10 \log_{10} \frac{255^2}{\left(\frac{1}{MN}\right) \sum_{i=0}^{M-1} \sum_{j=0}^{N-1} [SI(i,j) - CI(i,j)]^2} \tag{3}$$

where M and N are the number of rows and columns in the images, respectively. Higher values for the PSNR are preferable as they indicate better correspondence between the original image *CI* and the stego image *SI*. The pure embedded capacity is basically the total number of bits embedded in the stego image.

Many experiments were performed to evaluate the performance of the proposed algorithm and its extensions. The experiments include the performance evaluation of different algorithms at maximum embedding capacity and variable payload. Additionally, the experiments investigated the effect of using more than two predictors in the proposed algorithms. The details of these experiments are presented in the following subsections.

## 3.1. Evaluation under Maximum Embedding Capacity

The experiment here tests the performance of the proposed algorithm and its extensions and compares to other algorithms in terms of the maximum embedding capacity and the associated image quality. Table 1 and Table 2 list the pure embedding capacity and PSNR values for different algorithms, respectively.

For the 1-Bin MPE2 algorithm, it is clear from the numbers that it outperforms the original MPE [15], Hong, *et al*. [16], Li, *et al.* [17] and Lin, *et al.* [18] algorithms in terms of the maximum embedding capacity (EC). As a matter of fact, the 1-Bin proposed algorithm achieved a remarkable increase in the embedding capacity despite the fact that it uses one bin only for embedding. This is related to the fact that the proposed algorithm uses two predictors instead of one as in the 2-Bin MPE, which increases the possibility of obtaining more zero values for the prediction errors that are used for embedding.

Table 2 lists the PSNR values of different algorithms. It is obvious how the PSNR values of the 1-Bin MPE2 are slightly higher than those of the 2-Bin MPE. This related to the fact that some of the pixels in our algorithm are not modified when the prediction errors are bipolar. When the PSNR values of Hong, *et al.* [16], Li, *et al.* [17] and Lin, *et al.* [18] are compared with those of the 1-Bin MPE2, it is clear that the1-Bin MPE has relatively lower values. However, this little difference is hardly noticeable when the stego images from different algorithms are compared visually as shown in Figure 7. This makes the proposed algorithm highly competitive as it has higher embedding capacity with comparable visual quality for the stego images. The following simple theoretical analysis verifies the PSNR values obtained in the 1-Bin proposed algorithm. In the worst case, assume that the probability of obtaining one out of the three cases for prediction errors (all positive, all negative and bipolar) encountered during embedding is 1/3 and all of the bits in the secret message are ones. Since all bits to embed are 1, prediction errors are incremented by 1 or decremented by 1 when all errors are positive and negative, respectively. However, no the errors are not modified in case of bipolar error. Accordingly, the PSNR value will be $10 \times \log_{10}(255/ (\frac{1^2+1^2}{3}))$ or 49.89 dB. However, the quality of stego image will be around this value since probability of obtaining bipolar errors is not always equal to 1/3 in addition to the fact that the secret message is usually a mix of zeros and ones.

Table 1. Comparison between different algorithms in terms of pure maximum embedding capacity (bits)

| Image | 2-Bin MPE [15] | 3-Bin MPE | Hong *et al.* [16] | Li, *et al.* [17] | Lin, *et al.* [18] | The proposed algorithm (MPE2) | | |
| --- | --- | --- | --- | --- | --- | --- | --- | --- |
| | | | | | | 1-Bin | 2-Bin | 3-Bin |
| **Lena** | 46,667 | 63,489 | 45,416 | 27,006 | 33,111 | 57,406 | 77,746 | 101,121 |
| **Baboon** | 39,863 | 56,263 | 13,024 | 15,665 | 18,224 | 40,957 | 48,250 | 64,019 |
| **Airplane** | 66,153 | 82,522 | 62,813 | 36,193 | 46,367 | 69,213 | 89,184 | 116,517 |
| **Peppers** | 60,160 | 79,803 | 35,603 | 35,688 | 44,922 | 62,796 | 82,419 | 106,634 |
| **Boat** | 45,705 | 62,174 | 28,739 | 30,711 | 37,367 | 46,760 | 64,211 | 83,615 |
| **Barbara** | 35,727 | 49,268 | 34,423 | 21,361 | 25,669 | 37,685 | 50,445 | 65,619 |
| **Average** | **49,046** | **65,587** | **36,670** | **27,771** | **34,277** | **52,470** | **68,709** | **89,588** |

Table 2. Comparison between different algorithms in terms of PSNR (dB) at maximum capacity

| Image | 2-Bin MPE [15] | 3-Bin MPE | Hong *et al.* [16] | Li, *et al.* [17] | Lin, *et al.* [18] | The proposed algorithm (MPE2) | | |
| --- | --- | --- | --- | --- | --- | --- | --- | --- |
| | | | | | | 1-Bin | 2-Bin | 3-Bin |
| **Lena** | 48.56 | 44.95 | 49.98 | 52.31 | 50.91 | 49.64 | 46.84 | 44.79 |
| **Baboon** | 48.49 | 44.91 | 49.69 | 51.77 | 50.42 | 49.67 | 46.36 | 44.44 |
| **Airplane** | 48.73 | 45.20 | 50.15 | 53.08 | 51.61 | 49.78 | 47.16 | 45.14 |
| **Peppers** | 48.69 | 45.21 | 50.05 | 52.85 | 51.45 | 49.74 | 47.03 | 45.01 |
| **Boat** | 48.55 | 44.96 | 49.90 | 52.53 | 51.07 | 49.65 | 46.60 | 44.60 |
| **Barbara** | 48.45 | 44.79 | 49.79 | 52.03 | 50.65 | 49.52 | 46.26 | 44.24 |
| **Average** | **48.58** | **45.00** | **49.93** | **52.43** | **51.02** | **49.67** | **46.71** | **44.70** |

Considering the 2-Bin MPE2 algorithm, Table 1 shows the significant increase in the embedding capacity due to incorporating one more bin in the embedding process. Comparing with the 2-Bin MPE shows the advantage of using two predictors in increasing the embedding capacity, despite the fact that these two algorithms use two bins for embedding. Nonetheless, this increase in embedding capacity comes at the cost of decreasing the PSNR values in the 2- Bin MPE2 algorithm. However, this decrease is hardly noticeable by the human eye since the PSNR values are higher than 40 dB [28] and as shown in Figure 7. Similar observations in terms of increasing the embedding capacity and the associated decrease in the PSNR values can be drawn when the 3-Bin MPE2 is considered and compared with other algorithms, especially the 3-Bin MPE algorithm. In this case, the increase in the embedding capacity and the comparable PSNR values for the 3-Bin MPE2 are related to using two predictors.

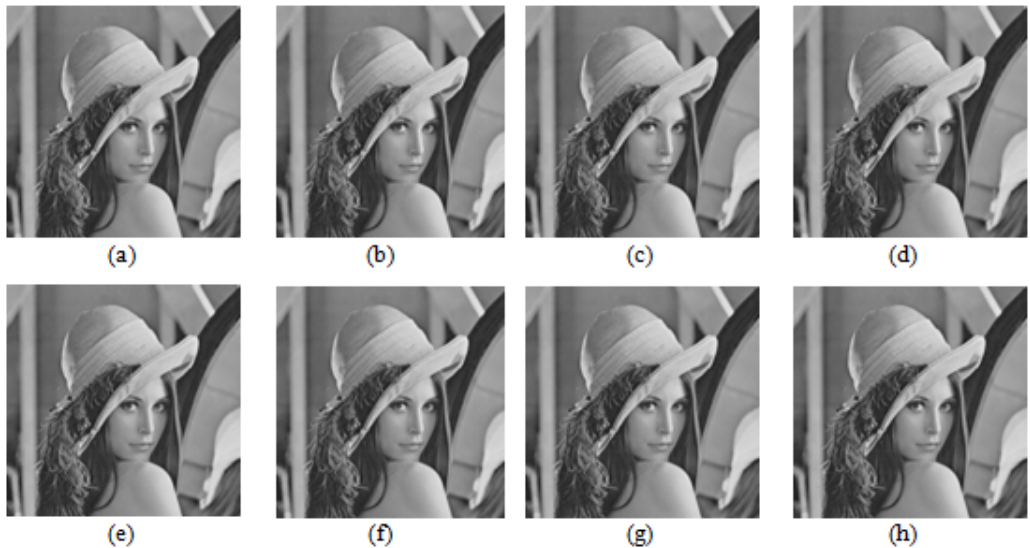

Figure 7. Results for different algorithms at maximum embedding capacity for the original image Lena
(a) 2-Bin MPE [15] (b) 3-Bin MPE (c) Hong *et al.* [16] (d) Li *et al.*[17] (e) Lin *et al.* [18]
(f) 1-Bin MPE2 (g) 2-Bin MPE2 (h) 3-Bin MPE2.

The theoretical analysis for 2-Bin version on the lower PSNR bound is similar to the discussion presented for the 1-Bin MPE2 algorithm except that negative prediction errors are decremented by 2 instead of 1 when the bit to be embedded is 1. Thus, the PSNR will be $10 \times \log_{10}(255/(\frac{1^2+2^2}{3}))$ or 45.91 dB. Similarly, the lower PSNR bound for the 3-Bin MPE2 algorithm will be $10 \times \log_{10}(255/(\frac{2^2+2^2}{3}))$ or 43.87 dB.

In addition to the results presented for the images shown in Figure 6, the proposed algorithm and its extensions were tested on 1000 images of size 512×512 selected from a large image database (Image Database Website) [29]. The maximum EC and PSNR values are listed in Table 3. The results shown in Table 3 reveal that the 3-Bin MPE2 algorithm performs better than all others algorithms in terms of embedding capacity However, and as discussed earlier, the increase in embedding capacity comes at the cost of lowering the PSNR values.  Additionally, the maximum payload of the 2-Bin version of the proposed algorithm is also higher than all others algorithms. Comparing the 2-Bin version to the 2-Bin MPE shows the power of the proposed algorithm. When higher PSNR values are considered, the 1-Bin version of the proposed algorithm achieves competitive values when compared with the original MPE [15], Hong, *et al*. [16], Li, *et al*. [17] and Lin, *et al*. [18] algorithms, but with higher EC.

### 3.2. Evaluation under Variable Payload

The proposed algorithm and its variants were also investigated when the payload is less than the maximum embedding capacity. Specifically, image quality in terms of PSNR values is assessed when small payloads are embedded in the images.  Figure 8 shows examples for the PSNR values of the different algorithms with variable payload size when applied to Lena, Cameraman, Baboon and Toys images. The figure clearly illustrates that 1-Bin MPE2 algorithm outperforms the original MPE algorithm [15] and its extensions. This is mainly due to the fact that some pixels in the cover image are not modified when the prediction errors are bipolar in the proposed 1-Bin MPE2 algorithm. Additionally, the use of two predictors improves the prediction accuracy. This implies that fewer pixels need to be modified to embed a specific payload, which improves the PSNR values. Comparing the proposed algorithm with Hong, *et al.* [16], Li, *et al.* [17] and Lin, *et al.* [18] algorithms, it is evident, for some images, how these algorithms outperform the proposed algorithm and its extensions. For example, in the Cameraman image, these algorithms have better PSNR values when the payload is lower than 40000 bits. However, the proposed 1-Bin algorithm performs better when the payload is increased.

Table 3. Comparison of average maximum embedding capacity (bits) and
PSNR (dB) on  images selected from a large image database [29]

| Algorithm | Maximum EC | PSNR |
|---|---|---|
| 2-Bin MPE (Hong, et al., 2009) | 61,983 | 48.74 |
| 3-Bin MPE | 73,765 | 45.23 |
| Hong, et al., (2011) | 64,777 | 50.19 |
| Li, et al., (2013) | 31,585 | 53,28 |
| Lin, et al., (2014) | 39,460 | 51,43 |
| 1-Bin MPE2 | 67,266 | 49.84 |
| 2-Bin MPE2 | 85,238 | 47.00 |
| 3-Bin MPE2 | 106,842 | 45.12 |

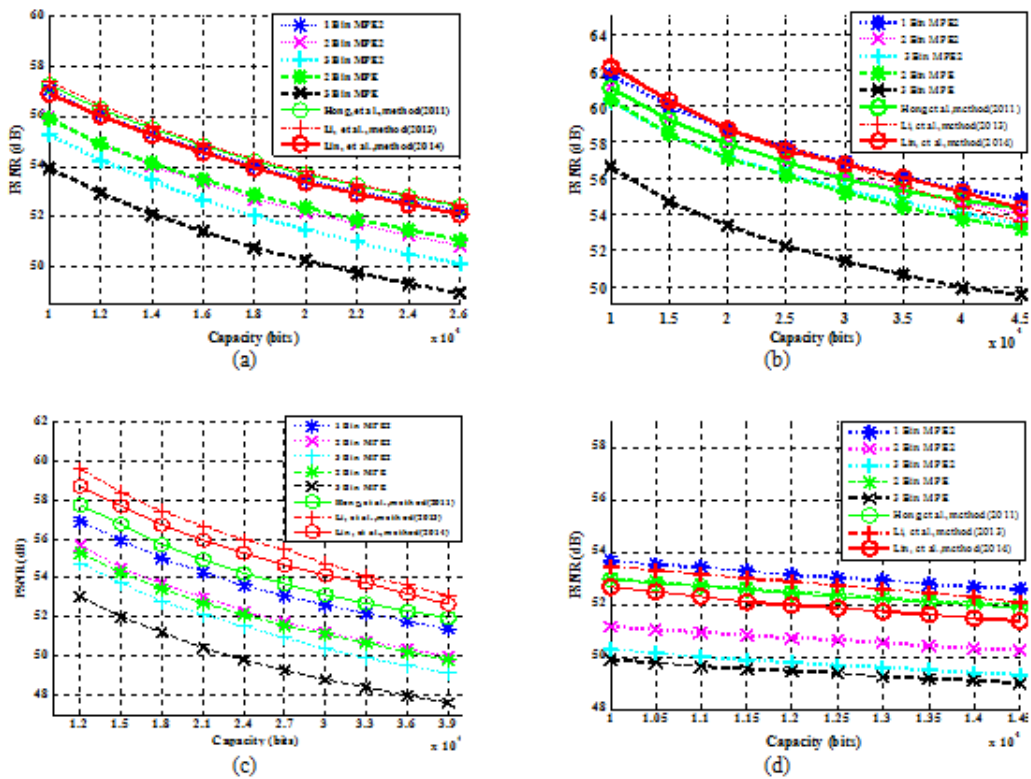

Figure 8. PSNR values for different algorithms under variable embedding capacity for  images
(a) Lena (b) Cameraman (c) Toys (d) Baboon.

Going back to Figure 8, it is obvious how the 2-Bin version and the 3-Bin version of the proposed algorithm outperform the 3-Bin version of the MPE algorithm in terms of PSNR at the same payload. The 2-Bin version of the proposed algorithm can do better than the original MPE algorithm, e.g. for images Toys and Baboon, even that 2-Bin version pixels may be shifted by 2 and in original MPE algorithm, the shift only by one. This shows the power of the proposed algorithm by utilizing the bins of two histograms of prediction error for embedding and when more accurate predictions is provided, it will result with modifying less pixels to embed the same payload as the image is scanned.

On overall, these results prove the efficiency of the proposed algorithms for embedding small and large payloads with competitive image quality. The 2-Bin version and 3-Bin version of the proposed algorithm have a lower PSNR values when compared to Li, *et al.* [17] and Lin, *et al.* [18] algorithms. This can be explained by the fact that the 2-Bin version and 3-Bin version of the proposed algorithms shift some pixels by more than 1 depending on the outcome of the prediction errors, while the algorithms in Li, *et al.* [17] and Lin, *et al.* [18] are designed to shift pixels by 1 at maximum. However, the maximum embedding capacity of 2-Bin version and 3-Bin version of the proposed algorithms are much higher than that of Li, *et al.* [17] and Lin, *et al.* [18]. Additionally, this small difference in PSNR is hardly noticeable by the human eye as we discussed earlier.

### 3.3. Evaluation with More than Two Predictors

It is known that the embedding capacity in prediction-based RDH algorithms depends on the prediction accuracy. Accordingly, if the number of predictors increases, the chance of obtaining prediction errors values that are used for embedding will increase. This would in turn result in increasing the embedding capacity, so; we considered the evaluation of the proposed algorithms when the number of predictors is increased. Specifically, we considered using three and four predictors. This evaluation requires no modification to the steps outlined in the embedding and extraction procedures. The only difference lies in the fact that more predictions are computed at each pixel in the image. The two additional predictors that we used in this evaluation are the median and minimum predictors. The Median and minimum of three neighbouring pixels computes the prediction value P using equation (4) and (5). The median of a set of data is the middle value of a sorted list [21]. The average of two middles is calculated if the number of list n is even. However, the median will be the value of number $n+1/2$ if the number of list n is odd.

$$P = \text{Median}([a\ b\ c]) \tag{4}$$

$$P = \text{Minimum}([a\ b\ c]) \tag{5}$$

Table 4 shows the embedding capacity and the PSNR values for the 1-Bin MPE2 algorithm and its extensions when two, three and four predictors are used. The numbers show a slight difference in the performance of the three algorithms when the number of predictors is increased. Although this contradicts with the general understanding that increasing the number of predictors should increase the prediction accuracy, the behaviour of the three algorithms can be simply explained by considering Figure 9 which shows the distribution of prediction errors in terms of polarity for four different images as the number of predictors is increased. The figure clearly shows how increasing the predictors reduce the possibility of obtaining unipolar errors which are used for embedding in the proposed algorithms. It is worth to mention that the increase in the PSNR values in some cases is related to the fact that using more predictors results in having more bipolar errors. This implies that the corresponding pixels are not modified according to the proposed algorithms.

Table 4. Comparison between different algorithms using different number of predictors in terms of pure maximum embedding capacity (bits) and PSNR

| Image | 1-Bin MPE2 Algorithm Using Two Predictors | | 1-Bin MPE3 Algorithm Using Three Predictors | | 1-Bin MPE4 Algorithm Using Four Predictors | |
|---|---|---|---|---|---|---|
| | Capacity | PSNR | Capacity | PSNR | Capacity | PSNR |
| **Lena** | 57,406 | 49.64 | 57,232 | 49.99 | 53,580 | 50.70 |
| **Baboon** | 38,957 | 49.67 | 31,151 | 49.99 | 27,199 | 50.65 |
| **Airplane** | 69,213 | 49.78 | 70,501 | 50.19 | 64,922 | 50.85 |
| **Peppers** | 60,796 | 49.74 | 60,553 | 50.11 | 55,702 | 50.84 |
| **Boat** | 46,760 | 49.65 | 46,764 | 49.97 | 44,025 | 50.66 |
| **Barbara** | 35,685 | 49.52 | 35,403 | 49.85 | 33,221 | 50.53 |

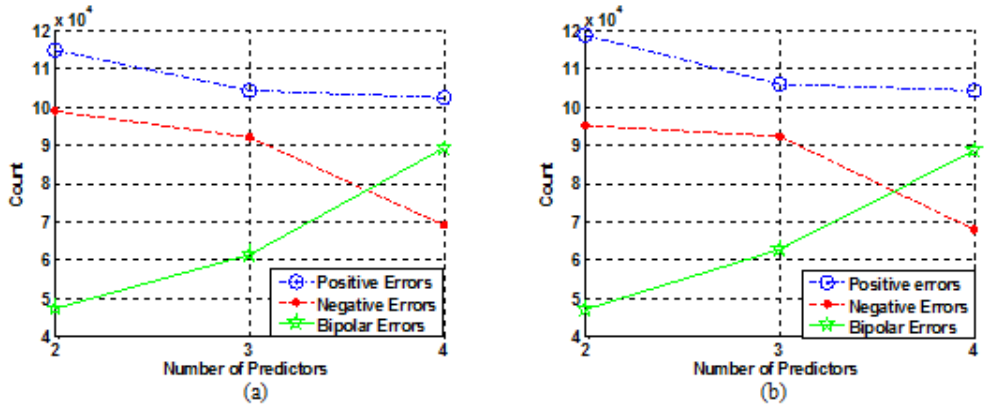

Figure 9. Distribution of prediction errors using variable number of predictors for images
(a) Lena (b) Airplane.

## 4. CONCLUSIONS

The goal of this paper is to improve the efficiency of prediction-based reversible data hiding algorithms by designing an algorithm that employs two predictors to improve the prediction accuracy, thus the embedding capacity. The proposed algorithm is based on the efficient modification of prediction errors (MPE) algorithm; however, it incorporates two predictors and uses only one bin of the prediction errors histogram for embedding the data, and it is referred to as 1-Bin MPE2. Performance evaluation of the proposed algorithm showed its ability to increase the embedding capacity with competitive image quality. Additionally, no overhead information is added to cope with the increase in the number of predictors. These results motivated us to propose two simple extensions extend the proposed algorithm. Basically, these extensions consider the use of more bins of the prediction errors histogram. Specifically, the original algorithm is modified to use two and three bins. These extensions are referred to by 2-Bin MPE2 and 3-Bin MPE2. These extensions provide significant increase in the embedding capacity with reasonable and acceptable image quality.

## AUTHORS

**Enas N. Jaara** received her B.Sc. and M.Sc. degrees in computer engineering from the University of Jordan in 2007 and 2015, respectively. She is currently working as a teaching assistant and lab supervisor in the Computer Engineering Department at the University of Jordan. Her research interests are in embedded systems and image processing.

**Iyad F. Jafar** received his Ph.D. degree in Computer Engineering from Wayne State University, Michigan, in 2008. He is currently working as an associate professor in the Department of Computer Engineering at the University of Jordan. His research interests are in signal and image processing, pattern recognition, multimedia and computer networks.